\documentstyle[12pt]{article}

\begin{document}

%%%%%%%%%%%%%%%%%%%%%%%%%%%%%%%%%%%%%%%%%%%%%%%%%%%%%%%%%%%%%%%%%%%%
%%%%%%%%%%%%%%%%%%%%%%%%%%%%%%  Defs. %%%%%%%%%%%%%%%%%%%%%%%%%%%%%%
%%%%%%%%%%%%%%%%%%%%%%%%%%%%%%%%%%%%%%%%%%%%%%%%%%%%%%%%%%%%%%%%%%%%

\def\a{\alpha}
\def\b{\beta}
\def\c{\varepsilon}
\def\d{\delta}
\def\e{\epsilon}
\def\f{\phi}
\def\g{\gamma}
\def\h{\theta}
\def\k{\kappa}
\def\l{\lambda}
\def\m{\mu}
\def\n{\nu}
\def\p{\psi}
\def\q{\partial}
\def\r{\rho}
\def\s{\sigma}
\def\t{\tau}
\def\u{\upsilon}
\def\v{\varphi}
\def\w{\omega}
\def\x{\xi}
\def\y{\eta}
\def\z{\zeta}
\def\D{\Delta}
\def\G{\Gamma}
\def\L{\Lambda}
\def\F{\Phi}
\def\P{\Psi}
\def\S{\Sigma}

\def\o{\over}

\def\IJMP{Int.~J.~Mod.~Phys. }
\def\MPL{Mod.~Phys.~Lett. }
\def\NP{Nucl.~Phys. }
\def\PL{Phys.~Lett. }
\def\PR{Phys.~Rev. }
\def\PRL{Phys.~Rev.~Lett. }
\def\PTP{Prog.~Theor.~Phys. }
\def\ZP{Z.~Phys. }

\def\beq{\begin{equation}}
\def\eeq{\end{equation}}

%%%%%%%%%%%%%%%%%%%%%%%%%%%%%%%%%%%%%%%%%%%%%%%%%%%%%%%%%%%%%%%%%%%%

\title{
\begin{flushright}
\large UT-792
\end{flushright}
       $R$-Invariant Unification with Dynamical Higgs Multiplets}

\author{K.-I. Izawa and T. Yanagida \\
        \\ {\sl Department of Physics, University of Tokyo, Tokyo 113, Japan}}

\date{October 6, 1997}

\maketitle\thispagestyle{empty}
\setlength{\baselineskip}{3.6ex}

\begin{abstract}
We construct $R$-invariant grand unification
models with composite Higgs multiplets.
Higgs doublets are massless due to the $R$ symmetry,
which is not spontaneously broken at the unification scale,
while Higgs triplets attain large masses from a dynamically generated
superpotential, which is allowed by an $R$ anomaly.
\end{abstract}

\newpage

\section{Introduction}

Fine tuning of coupling constants is required
when some terms in the Lagrangian are unnaturally small
without the possibility of a symmetry-based explanation.
In other words, the symmetries that could forbid such small terms
would inevitably also forbid some necessary terms.

There may be various ways to avoid this unnatural fine tuning.
One possible way is to utilize anomalous symmetries
which forbid small terms to be fine tuned
and whose quantum breaking may generate necessary terms.

The doublet-triplet splitting
in the supersymmetric grand unified theory (SUSY GUT)
\cite{Dim}
poses one of the most serious fine-tuning problems in particle physics.%
\footnote{There have been several attempts to solve this problem.
In particular, the semisimple unification approach
\cite{Yan,Hot,Iza}
provides completely natural models with all the successes
of the minimal SUSY GUT kept intact.
This motivates the present construction of composite Higgs models.}
In this paper,
we construct $R$-invariant unification models
with composite Higgs multiplets.
Higgs doublets are massless due to the $R$ symmetry,
which is not spontaneously broken at the GUT scale
\cite{Iza},
while Higgs triplets attain large masses from a dynamically generated
superpotential, which is allowed by an $R$ anomaly.

\section{The Model}

We extend the minimal SU$(5)_{GUT}$ GUT to a semisimple gauge
theory in order to incorporate a strong interaction which produces composite
Higgs multiplets.

We consider an ${\rm SU}(5)_{GUT} \times {\rm SU}(5)_H$ model.
The quarks and leptons obey the usual transformation law
under the GUT group SU$(5)_{GUT}$, while they are all singlets of the
hypercolor gauge group SU$(5)_H$.
This hypercolor interaction plays the role of making composite Higgs
doublets in the SUSY standard model.

We introduce six pairs of hyperquarks $Q_\a^I$ and ${\bar Q}_I^\a$
($\a = 1, \cdots, 5; I = 1, \cdots, 6$) which transform as ${\bf 5}$
and ${\bf 5}^*$ under the hypercolor group SU$(5)_H$.
The first five pairs $Q_\a^i$ and ${\bar Q}_i^\a$ ($i = 1, \cdots, 5$)
belong to ${\bf 5}^*$
and ${\bf 5}$ of SU$(5)_{GUT}$, respectively, and the last pair $Q_\a^6$
and ${\bar Q}_6^\a$ are singlets of SU$(5)_{GUT}$.

We also introduce an adjoint representation $\S_i^j$
of SU$(5)_{GUT}$
and two singlets ${\bar X}$ and $X$ whose $R$ charges are two
\cite{Iza}.
The superpotential for $Q_\a^I$, ${\bar Q}_I^\a$,
$\S_i^j$, ${\bar X}$ and $X$ is given by
\beq
\begin{array}{l}
\displaystyle
W = \S_i^j (\l Q_\a^i {\bar Q}_j^\a
  - \l' Q_\a^i {\bar Q}_k^\a Q_\b^k {\bar Q}_j^\b)
\\
\noalign{\vskip 2ex}
\displaystyle
  \qquad \ + f {\bar X} \e^{\a\b\g\d\c} Q_\a^1Q_\b^2Q_\g^3Q_\d^4Q_\c^5
  + f' X \e_{\a\b\g\d\c} {\bar Q}^\a_1{\bar Q}^\b_2
    {\bar Q}^\g_3{\bar Q}^\d_4{\bar Q}^\c_5,  
\end{array}
\label{POT}
\eeq
where we have assumed the vanishing $R$ charges
of $Q_\a^I$ and ${\bar Q}^\a_I$ and have omitted higher-order terms
which are not important for our purposes.
We have also imposed a U$(1)_X$ symmetry with the charges of
$Q_\a^I$, ${\bar Q}^\a_I$, ${\bar X}$ and $X$ as $1$, $-1$, $-5$ and $5$,
respectively.
Here we suppose that the higher-order terms are suppressed by
a cutoff (say the Planck) scale in the present effective theory
below the cutoff scale, which we set to unity.

\section{The Dynamics}

In the following sections, we investigate the model
by means of effective field theory analysis. 

The effective superpotential of the model may be written as
\beq
W_{eff} = \S_i^j (\l M^i_j - \l' M^i_k M^k_j)
        + f {\bar X} B_6 + f' X {\bar B}^6
        + \L^{-9}(B_I M^I_J {\bar B}^J - \det M^I_J),
\label{ESP}
\eeq
in terms of the chiral superfields
\beq
\begin{array}{l}
\displaystyle
M^I_J \sim Q_\a^I {\bar Q}_J^\a, \quad
B_I \sim \e_{IJKLMN}\e^{\a\b\g\d\c}Q_\a^JQ_\b^KQ_\g^LQ_\d^MQ_\c^N,
\\
\noalign{\vskip 2ex}
\displaystyle
{\bar B}^I \sim \e^{IJKLMN}\e_{\a\b\g\d\c}{\bar Q}^\a_J{\bar Q}^\b_K
                {\bar Q}^\g_L{\bar Q}^\d_M{\bar Q}^\c_N,
\end{array}
\eeq
where $\L$ denotes a dynamical scale of the SU$(5)_H$ interaction
\cite{Kap}.

The $F$-flatness condition yields vacuum expectation values satisfying
\beq
\begin{array}{l}
\displaystyle
\l M^i_j - \l' M^i_k M^k_j
 - {1 \o 5} \d^i_j (\l M^l_l - \l' M^l_k M^k_l) = 0,
\\
\noalign{\vskip 2ex}
\displaystyle
f{\bar X} + \L^{-9}M^6_J{\bar B}^J = 0, \quad f'X + \L^{-9}B_IM^I_6 = 0,
\\
\noalign{\vskip 2ex}
\displaystyle
B_6 = 0, \quad {\bar B}^6 = 0, \quad \det M^i_j = 0,
\end{array}
\label{FFL}
\eeq
where the last equation is a part of the condition
$\q W_{eff} / \q M^I_J = 0$.
Together with the $D$-flatness condition of the SU$(5)_{GUT}$ interaction
\cite{Dra},
we obtain the desired form
\cite{Wil}
of the vacuum expectation value of $M^i_j$
with $\S_i^j = 0$, ${\bar X} = 0$, $X = 0$,
$B_i = 0$, ${\bar B}^i = 0$, $M_i^6 = 0$ and $M_6^i = 0$:
\beq
  \begin{array}{c}
    \displaystyle M^i_j = {\l \o \l'}
    \left(
      \begin{array}{ccccc}
        1 & & & & \\
        & 1 & & & \\
        & & 1 & & \\
        & & & 0 & \\
        & & & & 0
      \end{array}
    \right), \\
  \end{array}
\label{VEV}
\end{equation}
where the vanishing eigenvalues%
\footnote{Another form of the $M^i_j$ with three vanishing eigenvalues
is possible. It provides massless color triplets at low energy.
This may be utilized to produce massless weak doublets through the
missing partner mechanism
(see Ref.\cite{Yan,Hot,Iza}).}
are enforced by the condition $\det M^i_j = 0$ in Eq.({\ref{FFL}).
The GUT scale is expected to be of order $\l / (\l' \L)$.
We note that the $R$ symmetry is unbroken in this vacuum
\cite{Iza}.

\section{Low-Energy Spectrum}

Let us consider the low-energy mass spectrum of the model
in the vacuum Eq.(\ref{VEV}). 

The color triplets in the composites $B_i$ and ${\bar B}^i$ acquire
large masses%
\footnote{The mass may be larger than the GUT scale
due to the strong-coupling nature of the hypercolor interaction
\cite{Lut}.
This suppresses the Higgsino-mediated proton decay
through the dimension-five operator
\cite{Sak}.
We note that the dimension-five operators are severely
suppressed in the semisimple unification
\cite{Yan},
except for the model in Ref.\cite{Hot}.
}
due to the form Eq.(\ref{VEV}) of the vacuum expectation value of $M^i_j$
from the interaction term $B_I M^I_J {\bar B}^J$
in the effective superpotential Eq.({\ref{ESP}).

On the other hand, the weak doublets $B_i$ and ${\bar B}^i$ ($i = 4, 5$)
remain massless due to the vanishing of the eigenvalues
of $M^i_j$ in Eq.(\ref{VEV}). 
We may regard these doublets as those in the SUSY standard model
and introduce their Yukawa interactions with quarks and leptons.%
\footnote{The U$(1)_X$ charges of ${\bf 10}$ and ${\bf 5}^*$ of SU$(5)_{GUT}$
are $5/2$ and $-15/2$, respectively.
U$(1)_X$ may be regarded as a gauge group when we introduce
singlets which cancel the gauge anomaly.
These singlets may be regarded as right-handed neutrinos.}
We are led to consider the case that the dynamical scale $\L$ is 
of order 1,%
\footnote{If $\L \ll 1$, the masses of some components in the adjoints
multiplets of SU$(5)_{GUT}$ would be smaller than the GUT scale.
This would raise the GUT scale \cite{Bac}.} 
so as to have the heavy top quark with a large Yukawa coupling.

In addition, $M_i^6$ and $M^i_6$ yield massless complete multiplets%
\footnote{They never have Yukawa couplings with the quarks and leptons,
owing to the U$(1)_X$ symmetry.}
of SU$(5)_{GUT}$, and $M_6^6$ gives a massless singlet.
They are protected from having masses by the $R$ symmetry,
as is the case for the composite Higgs doublets
$B_i$ and ${\bar B}^i$ ($i = 4, 5$).
The $R$ (and SUSY) breaking may
allow these multiplets to have masses on the order of the weak scale.
This may stabilize our vacuum $M^6_i = M^i_6 = 0$.
\footnote{Note added: We notice that
an additional field with the $R$-charge $(-2)$
is necessary to give masses for
$(\bf 3,2^*)$ and $(\bf 3^*,2)$ 
components in $\Sigma_i^j$,
since $(\bf 3,2^*)$ and $(\bf 3^*,2)$ in $M^i_j$ 
are absorbed into broken gauge bosons.
This can be achieved along the lines of section 2
in our subsequent paper Phys.~Rev.~D60 (1999) 115016
with the successes of the present model intact.
}

\section{Conclusion}

We have constructed an $R$-invariant
${\rm SU}(5)_{GUT} \times {\rm SU}(5)_H$
unification model which, in particular, naturally
yields the SUSY standard model
with the weak doublets.
The SU$(5)_H$ hypercolor gauge interaction provides the weak doublets
as composites of the hyperquarks without its color triplet partners at
low energy. The triplets attain large masses from the dynamically
generated superpotential of the SU$(5)_H$ interaction,
which is allowed by an $R$ anomaly,
while the $R$ symmetry preserves the masslessness of the doublets
as long as it is unbroken together with the SUSY.
This model predicts,
in addition to the contents of the minimal SUSY standard model,
a pair of ${\bf 5}$ and ${\bf 5}^*$ of SU$(5)_{GUT}$
and a singlet at the weak scale.
These predictions may be tested in future collider experiments.

Along similar lines of construction,
we may consider unification models 
with dynamical Higgs multiplets
by means of other semisimple gauge groups.
For example, the composite Higgs multiplets in the SU$(5)_{GUT}$ GUT
may be predicted by an SU$(4)_H$ hypercolor gauge theory
with five flavors of hyperquarks instead of the SU$(5)_H$ theory.
Namely, we may consider an ${\rm SU}(5)_{GUT} \times {\rm SU}(4)_H$ model.
An extension to the SO$(10)_{GUT}$ GUT is also possible.
We may construct an ${\rm SO}(10)_{GUT} \times {\rm SU}(10)_H$ model
with eleven flavors of hyperquarks in a similar fashion to the
${\rm SU}(5)_{GUT} \times {\rm SU}(5)_H$ model.

\end{document}